\documentclass[pra,twocolumn,graphicx]{revtex4}
\usepackage{epsfig}
\usepackage{bm}
\usepackage{amsmath}

\begin{document}

\title{Quantum interference in laser-induced nonsequential double ionization
in diatomic molecules: the role of alignment and orbital symmetry}
\author{C. Figueira de Morisson Faria$^{1},$ T. Shaaran$^{1}$, X. Liu$^{2}$
and W. Yang$^{1,3}$}
\affiliation{$^1$Department of Physics and Astronomy, University College London, Gower
Street, London WC1E 6BT, United Kingdom\\
$^{2}$State Key Laboratory of Magnetic Resonance and Atomic and Molecular
Physics, Wuhan Institute of Physics and Mathematics, Chinese Academy of
Sciences, Wuhan 430071, China \\
$^{3}$State Key Laboratory of High-Field Laser Physics, Shanghai Institute
of Optics and Fine Mechanics, Chinese Academy of Sciences, Shanghai 201800,
China}
\date{\today}

\begin{abstract}
We address the influence of the orbital symmetry and of the molecular
alignment with respect to the laser-field polarization on laser-induced
nonsequential double ionization of diatomic molecules, in the length and
velocity gauges. We work within the strong-field approximation and assume
that the second electron is dislodged by electron-impact ionization, and
also consider the classical limit of this model. We show that the
electron-momentum distributions exhibit interference maxima and minima due
to the electron emission at spatially separated centers. The interference
patterns survive the integration over the transverse momenta for a small
range of alignment angles, and are sharpest for parallel-aligned molecules.
Due to the contributions of transverse-momentum components, these patterns
become less defined as the alignment angle increases, until they disappear
for perpendicular alignment. This behavior influences the shapes and the
peaks of the electron momentum distributions.
\end{abstract}

\maketitle

\address{$^1$Department of Physics and Astronomy, University College London,
Gower Street, London WC1E 6BT, United Kingdom\\
$^{2}$State Key Laboratory of Magnetic Resonance and Atomic and Molecular
Physics, Wuhan Institute of Physics and Mathematics, Chinese Academy of
Sciences, Wuhan 430071, China \\
$^{3}$State Key Laboratory of High-Field Laser Physics, Institute of Physics
and Fine Mechanics, Chinese Academy of Sciences, Shanghai 201800, China}

\section{Introduction}

High-order harmonic generation (HHG) or above-threshold ionization (ATI) owe
their existence to the recombination, or elastic collision, respectively, of
an electron with its parent ion or molecule \cite{tstep}. Such processes
take place within a fraction of a laser cycle, which, for a typical
Titanium-Sapphire, high-intensity laser pulse is $T\sim 2.7fs$. Therefore,
HHG and ATI occur within hundreds of attoseconds \cite{Scrinzi2006}. As a
direct consequence, one may use both phenomena to retrieve information from
a molecule, in particular the configuration of ions with which the active
electron recombines or rescatters, with subfemtosecond and subangstrom
precision. Concrete examples include the tomographic reconstruction of
molecular orbitals \cite{Itatani}, and the probing of structural changes in
molecules with attosecond precision \cite{attomol}.

In particular diatomic molecules, due to their simplicity, have attracted a
great deal of attention \cite%
{TDSE2005,doubleslit,KB2005,KBK98,MBBF00,Madsen,Usach2006,Usachenko,HenrikDipl,Kansas,PRACL2006,moreCL,DM2006,HBF2007,F2007}%
. For such systems, sharp interference maxima and minima in the ATI or HHG
spectra have been identified, which could be explained by the fact that
harmonic or photoelectron emission took place at spatially separated sites.
Therefore, a diatomic molecule may be viewed as the microscopic counterpart
of a double-slit experiment \cite{doubleslit}. 

Apart from the above-mentioned phenomena, nonsequential double or multiple
ionization, may, in principle, also be employed to extract information about
the structure of a molecule. This is not surprising, since laser-induced
recollision also plays an important role in this case. The main difference
from the previous scenarios is that the returning electron rescatters
inelastically with its parent ion, or molecule, giving part of its kinetic
energy to release other electrons. In particular, for the simplest case of
nonsequential double ionization (NSDI), this may occur through direct,
electron-impact ionization, or by excitation-tunneling mechanisms, in which,
upon recollision, the first electron excites the second. Thereafter, the
second electron tunnels out from an excited state. If more than two
electrons are involved, there may exist more complex ionization mechanisms
subsequent to recollision. Recently, a statistical thermalization model has
been proposed in order to account for these mechanisms and provide a simple
description of nonsequential multiple ionization \cite{therm2006}.

Specifically for NSDI of molecules, early measurements have already revealed
that the total double ionization yields depend on the molecular species, and
that there is an intensity region for which the predictions of sequential
models break down \cite{Cor98}. Furthermore, recent NSDI experiments on
diatomic molecules have shown that the shapes of the electron momentum
distributions depend on the symmetry of the highest occupied molecular
orbital. This holds even if the molecular sample is randomly aligned with
respect to the laser-field polarization \cite{NSDIsymm}. Indeed, in \cite%
{NSDIsymm}, very distinct electron momentum distributions have been observed
for $N_{2}$ and $\ O_{2}$, as functions of the electron momentum components $%
p_{n\parallel }$ $(n=1,2)$ parallel to the laser-field polarization. For the
former species, elongated maxima along the diagonal $p_{1\parallel
}=p_{2\parallel }$ have been reported, while, for $O_{2},$ the distributions
exhibit a prominent maximum in the region of vanishing parallel momenta, and
are quite broad along the direction $p_{1\parallel }=-p_{2\parallel }+const$%
. This has been confirmed by theoretical computations within a classical
framework, which reproduced some of the differences in the yields.

Subsequently, it has been found that the peak momenta and the shape of the $%
N_{2}$ electron-momentum distributions changed considerably with the
alignment angle of the molecules, with respect to the laser-field
polarization \cite{NSDIalign}. Specifically, for parallel alignment, roughly
$40\%$ larger peak momenta along the diagonal $p_{1\parallel }=p_{2\parallel
}$ have been observed, as compared to the perpendicular case. Furthermore,
for perpendicular alignment, a larger number of events in the second and
fourth quadrant of the momentum plane $(p_{1\parallel },p_{2\parallel })$
has been reported. In \cite{NSDIalign}, these events have been attributed to
excitation-tunneling mechanisms.

Despite the above-mentioned investigations, NSDI in molecules has been
considerably less studied than HHG or ATI, possibly due to the fact that it
is far more difficult to measure, or to model \cite{NSDIreview}. Solely from
a theoretical viewpoint, even for a single atom, it is a very demanding task
to perform a fully numerical, three-dimensional computation of NSDI
differential electron momentum distributions, for the parameter range of
interest. Indeed, only very recently, this has been achieved for Helium \cite%
{RuizBeck}, which is the simplest species for which NSDI occurs. An
alternative to that are semi-analytical approaches, mostly within the
framework of the strong-field approximation (SFA). Such methods are easier
to implement and provide a more transparent physical interpretation of the
problem. They exhibit, however, several contradictions. A concrete example
is the fact that cruder types of electron-electron interaction, such as a
contact-type interaction at the position of the ion, lead to a better
agreement with the experiments as compared to more refined choices, such as
a long-range Coulomb-type interaction \cite{FSLB2004,FL2005}. Apart from
that, it is not clear whether the interference patterns due to electron
emission in spatially separated centers survive the integration over the
transverse momentum components. This is particularly important, as in most
NSDI experiments, these quantities are not resolved.

These issues add up to the already vast amount of open questions related to
HHG and ATI in molecules. Indeed, even for these considerably more studied
phenomena, topics such as the gauge dependence of the interference patterns
\cite{PRACL2006,DM2006,F2007}, the role of different scattering or
recombination scenarios \cite{KBK98,Usach2006,PRACL2006,HBF2007,F2007}, and
the influence of collective effects \cite{collective} or polarization \cite%
{DM2006,dressedSFA,BCM2007} on the electronic bound states have raised
considerable debate. Apart from that, specifically for NSDI, it is not even
clear if the interference patterns would survive the integration over the
transverse electron momenta. This is particularly important, as, in most
NSDI experiments, the transverse momenta of the two electrons is not
resolved.

In this paper, we perform a systematic analysis of quantum-interference
effects in NSDI of diatomic molecules. We work within the strong-field
approximation (SFA) and assume that the second electron is dislodged by
electron-impact ionization. We also employ the classical limit of this
model. We consider frozen nuclei, and the linear combination of atomic
orbitals (LCAO approximation). This very simplified model has the main
advantage of allowing a transparent picture of the physical mechanisms
behind the interference patterns. Specifically, we investigate the influence
of the orbital symmetry and of the alignment angle on the NSDI electron
momentum distributions, and whether, within our framework, the features
reported in \cite{NSDIsymm} and \cite{NSDIalign} are observed. Furthermore,
we address the question of whether well-defined interference patterns such
as those observed in ATI or HHG computations may also be obtained for NSDI,
and, if so, under which conditions.

This paper is organized as follows. In Sec. \ref{transampl} we briefly
recall the expression for the NSDI transition amplitude, which is
subsequently applied to diatomic molecules (Sec. \ref{prefactors}).
Thereafter, we employ this approach to compute differential electron
momentum distributions, for angle-integrated (Sec. \ref{gauge}), and aligned
molecules (Sec. \ref{orbits}). In the latter case, we analyze the dependence
of the interference patterns on the alignment conditions. Finally, in Sec. %
\ref{concl}, we state the main conclusions of this paper.

\section{Transition amplitudes}

\subsection{General expressions}

\label{transampl} The simplest process behind NSDI corresponds to the
scenario in which the first electron, upon return, releases the second by
electron-impact ionization. The pertaining transition amplitude, within the
SFA, reads%
\begin{equation}
M(\mathbf{p}_{n},t,t^{\prime })=\int_{-\infty }^{\infty }\hspace*{-0.3cm}dt%
\hspace*{-0.1cm}\int_{-\infty }^{t}\hspace*{-0.3cm}dt^{\prime }\hspace*{%
-0.1cm}\int d^{3}kV_{p_{n}k}V_{k0}e ^{iS(\mathbf{p}_{n},\mathbf{k}%
,t,t^{\prime })},  \label{Mp}
\end{equation}%
with the action%
\begin{eqnarray}
S(\mathbf{p}_{n},\mathbf{k},t,t^{\prime })
&=&-\sum_{n=1}^{2}\int_{t}^{\infty }\hspace{-0.1cm}\frac{[\mathbf{p}_{n}+%
\mathbf{A}(\tau )]^{2}}{2}d\tau  \notag \\
&&-\int_{t^{\prime }}^{t}\hspace{-0.1cm}\frac{[\mathbf{k}+\mathbf{A}(\tau
)]^{2}}{2}d\tau -E_{02}t-E_{01}t^{\prime }  \label{singlecS}
\end{eqnarray}%
and the prefactors%
\begin{equation}
V_{\mathbf{k}0}=<\mathbf{\tilde{k}}(t^{\prime })|V|\phi _{0}^{(1)}>
\label{Vk0}
\end{equation}%
and%
\begin{equation}
V_{\mathbf{p}_{n}\mathbf{k}}=<\mathbf{\tilde{p}}_{1}\left( t\right) ,\mathbf{%
\tilde{p}}_{2}\left( t\right) |V_{12}|\mathbf{\tilde{k}}(t),\phi _{0}^{(2)}>.
\label{Vpnk}
\end{equation}%
Eq. (\ref{Mp}) describes the physical process in which an electron,
initially in a bound state $|\phi _{0}^{(1)}>,$ is released by tunneling
ionization at a time $t^{\prime }$ into a Volkov state $|\mathbf{\tilde{k}}%
(t)>$. Subsequently, this electron propagates in the continuum from $%
t^{\prime }$ to a later time $t.$ At this time, it is driven back by the
field and frees a second electron, which is bound at $|\phi _{0}^{(2)}>,$
through the interaction $V_{12}.$ Finally, both electrons are in Volkov
states. The final electron momenta are described by $\mathbf{p}_{n}(n=1,2).$
In the above-stated equations, $E_{0n}$ $(n=1,2)$ give the ionization
potentials, and $V$ the binding potential of the system in question. The
form factors (\ref{Vk0}) and (\ref{Vpnk}) contain all the information about
the binding potential, and the interaction by which the second electron is
dislodged, respectively.

Clearly, $V_{\mathbf{k}0}$ and $V_{\mathbf{p}_{n}\mathbf{k}}$ are gauge
dependent. In fact, in the length gauge $\mathbf{\tilde{p}}_{n}\left( \tau
\right) =\mathbf{p}_{n}+\mathbf{A}(\tau )$ and $\mathbf{\tilde{k}}(\tau )=%
\mathbf{k}+\mathbf{A}(\tau )(\tau =t,t^{\prime }),$ while in the velocity
gauge $\mathbf{\tilde{p}}_{n}\left( \tau \right) =\mathbf{p}_{n}$ and $%
\mathbf{\tilde{k}}(\tau )=\mathbf{k}.$ This is a direct consequence of the
fact that the gauge transformation $\chi _{l\rightarrow v}=\exp [-i\mathbf{A}%
(\tau )\cdot \mathbf{r}]$ from the length to the velocity gauge causes a
translation $\mathbf{p}\rightarrow \mathbf{p}-\mathbf{A}(\tau )$ in momentum
space. This is particularly critical for spatially extended systems, such as
molecules, and may alter the interference patterns. A similar effect has
also been investigated for high-order harmonic generation \cite%
{dressedSFA,F2007} and above-threshold ionization \cite{DM2006,BCM2007}.
These discrepancies can be overcome by adequately dressing the electronic
bound states, in order to make them gauge equivalent \cite{F2007}. In this
work, we will restrict ourselves to the field-undressed case. We expect,
however, based on the results of \cite{F2007}, that the field-dressed
velocity gauge distributions will have very similar patterns to the
field-undressed length gauge distributions. The same will hold for their
dressed length-gauge and the undressed velocity-gauge counterparts.

\subsection{Diatomic molecules}

\label{prefactors}

We will now consider the specific case of diatomic molecules. For
simplicity, we will assume frozen nuclei, the linear combination of atomic
orbitals (LCAO) approximation, and homonuclear molecules. Explicitly, the
molecular bound-state wave function for each electron reads
\begin{equation}
\psi _{0}^{(n)}(\mathbf{r}_{n})=C_{\psi }\left[ \phi _{0}^{(n)}(\mathbf{r}%
_{n}-\mathbf{R}/2)+\epsilon \phi _{0}^{(n)}(\mathbf{r}_{n}+\mathbf{R}/2)%
\right]
\end{equation}%
where $n=1,2,$ $\epsilon =\pm 1$, and $C_{\psi }=1/\sqrt{2(1+\epsilon S(%
\mathbf{R})}$, with
\begin{equation}
S(\mathbf{R})=\int \left[ \phi _{0}^{(n)}(\mathbf{r}_{n}-\mathbf{R}/2)\right]
^{\ast }\phi _{0}^{(n)}(\mathbf{r}_{n}+\mathbf{R}/2)d^{3}r.
\end{equation}%
\ The positive and negative signs for $\epsilon $ correspond to bonding and
antibonding orbitals, respectively. The binding potential of this molecule,
as seen by each electron, is given by
\begin{equation}
V(\mathbf{r}_{n})=V_{0}(\mathbf{r}_{n}-\mathbf{R}/2)+V_{0}(\mathbf{r}_{n}+%
\mathbf{R}/2),
\end{equation}%
where $V_{0}$ corresponds to the binding potential of each center in the
molecule.

The above-stated assumptions lead to
\begin{equation}
V_{\mathbf{k}0}^{(b)}=-\frac{2C_{\psi }}{(2\pi )^{3/2}}\cos [\mathbf{\tilde{k%
}}(t^{\prime })\cdot \mathbf{R}/2]\mathcal{I}(\mathbf{\tilde{k}}(t^{\prime
}))  \label{Vk0bond}
\end{equation}%
or%
\begin{equation}
V_{\mathbf{k}0}^{(a)}=-\frac{2iC_{\psi }}{(2\pi )^{3/2}}\sin [\mathbf{\tilde{%
k}}(t^{\prime })\cdot \mathbf{R}/2]\mathcal{I}(\mathbf{\tilde{k}}(t^{\prime
})),  \label{Vk0anti}
\end{equation}%
for the bonding and antibonding cases, respectively, with%
\begin{equation}
\mathcal{I}(\mathbf{\tilde{k}}(t^{\prime }))=\int d^{3}r_{1}\exp [i\mathbf{%
\tilde{k}}(t^{\prime })\cdot \mathbf{r}_{1}]V_{0}(\mathbf{r}_{1})\phi
_{0}^{(1)}(\mathbf{r}_{1}).
\end{equation}%
Thereby, we have neglected the integrals for which the binding potential $%
V_{0}(\mathbf{r})$ and the bound-state wave function $\phi _{0}^{(1)}(%
\mathbf{r})$ are localized at different centers in the molecule. We have
verified that the contributions from such integrals are very small for the
parameter range of interest, as they decrease very quickly with the
internuclear distance.

Eqs. (\ref{Vk0bond}) and (\ref{Vk0anti}) do not play a significant role in
the appearance of well-defined interference patterns. This is due to the
fact that the times $t^{\prime }$ at which the electron is emitted lie near
the peak field of the laser field. In other words, the electron trajectories
relevant to the momentum distributions start near the times for which the
electric field is maximum. For those most important trajectories, the range
of $k(t^{\prime })$ is so limited that the term $\cos (\mathbf{k}(t^{\prime
})\cdot \mathbf{R}/2)$ does not cross zero. In fact, we verified that the
prefactor $V_{\mathbf{k}0}$ has no influence on the interference patterns
(not shown).

Assuming that the electron-electron interaction depends only on the
difference between the coordinates of both electrons, i.e., $V_{12}=V_{12}(%
\mathbf{r}_{1}-\mathbf{r}_{2}),$ one may write the prefactor $V_{\mathbf{p}%
_{n}\mathbf{k}}$ as
\begin{equation}
V_{\mathbf{p}_{n}\mathbf{k}}^{(b)}=\frac{2C_{\psi }}{(2\pi )^{9/2}}V_{12}(%
\mathbf{p}_{1}-\mathbf{k})\cos [\mathbf{\mathcal{P}}(t)\cdot \mathbf{R}%
/2]\varphi _{0}^{(2)}(\mathbf{\mathcal{P}}(t))  \label{Vpnkbond}
\end{equation}%
or%
\begin{equation}
V_{\mathbf{p}_{n}\mathbf{k}}^{(a)}=\frac{2iC_{\psi }}{(2\pi )^{9/2}}V_{12}(%
\mathbf{p}_{1}-\mathbf{k})\sin [\mathbf{\mathcal{P}}(t)\cdot \mathbf{R}%
/2]\varphi _{0}^{(2)}(\mathbf{\mathcal{P}}(t)),  \label{Vpnkanti}
\end{equation}%
with $\mathbf{\mathcal{P}}(t)=\mathbf{\tilde{p}}_{1}(t)+\mathbf{\tilde{p}}%
_{2}(t)-\mathbf{\tilde{k}}(t)$, for bonding and antibonding orbitals,
respectively. Thereby,
\begin{equation}
\varphi _{0}^{(2)}(\mathbf{\mathcal{P}}(t))=\int d^{3}r_{2}\exp [i\mathbf{%
\mathbf{\mathcal{P}}}(t)\cdot \mathbf{r}_{2}]\phi _{0}^{(2)}(\mathbf{r}_{2}),
\end{equation}%
and

\begin{equation}
V_{12}(\mathbf{p}_{1}-\mathbf{k})=\int d^{3}rV_{12}(\mathbf{r})\exp [i(%
\mathbf{p}_{1}-\mathbf{k})\cdot \mathbf{r}],
\end{equation}%
with $\mathbf{r=r}_{1}-\mathbf{r}_{2}.$ Specifically, in the velocity and
length gauges, the argument in Eqs. (\ref{Vpnkbond}), (\ref{Vpnkanti}) is
given by $\mathbf{\mathcal{P}}(t)=\mathbf{p}_{1}+\mathbf{p}_{2}-\mathbf{k}$
and $\mathbf{\mathcal{P}}(t)=\mathbf{p}_{1}+\mathbf{p}_{2}-\mathbf{k+A}(t)$,
respectively.

The interference patterns studied in this work are caused by the pre-factors
$V_{\mathbf{p}_{n}\mathbf{k}}$. Explicitly, the two-center interference
condition defined by $V_{\mathbf{p}_{n}\mathbf{k}}$ gives the extrema
\begin{equation}
\left[ \mathbf{\tilde{p}}_{1}(t)+\mathbf{\tilde{p}}_{2}(t)-\mathbf{\tilde{k}}%
(t)\right] \cdot \mathbf{R}=n\pi .  \label{interference}
\end{equation}%
For symmetric highest occupied molecular orbitals, even and odd numbers in
Eq. (\ref{interference}) denote maxima and minima, respectively, whereas in
the antisymmetric case the situation is reversed (i.e., even and odd $n$
give minima and maxima, respectively). The above-stated equation will be
discussed in more detail in Sec. III.B.

The structure of the highest occupied molecular orbital is embedded in Eqs. (%
\ref{Vk0bond})-(\ref{Vpnkanti}). The simplest way to proceed is to consider
these prefactors and the single-center action (\ref{singlecS}). The multiple
integral in (\ref{Mp}) will be solved using saddle-point methods. For that
purpose, we must find the coordinates $(t_{s},t_{s}^{\prime },\mathbf{k}%
_{s}) $ for which $S(\mathbf{p}_{n},\mathbf{k},t,t^{\prime })$ is
stationary, i.e., for which the conditions $\partial _{t}S(\mathbf{p}_{n},%
\mathbf{k},t,t^{\prime })=\partial _{t^{\prime }}S(\mathbf{p}_{n},\mathbf{k}%
,t,t^{\prime })=0$ and $\partial _{\mathbf{k}}S(\mathbf{p}_{n},\mathbf{k}%
,t,t^{\prime })=\mathbf{0}$ are satisfied. This leads to the equations
\begin{equation}
\left[ \mathbf{k}+\mathbf{A}(t^{\prime })\right] ^{2}=-2E_{01},
\label{saddle1}
\end{equation}%
\begin{equation}
\int_{t^{\prime }}^{t}d\tau \left[ \mathbf{k}+\mathbf{A}(\tau )\right] =0,
\label{saddle2}
\end{equation}%
and
\begin{equation}
\sum_{n=1}^{2}\frac{[\mathbf{p}_{n}+\mathbf{A}(t)]^{2}}{2}=\frac{\left[
\mathbf{k}+\mathbf{A}(t)\right] ^{2}}{2}-E_{02}.  \label{saddle3}
\end{equation}%
Eq. (\ref{saddle1}) gives the conservation of energy at the time $\
t^{\prime },$ at which the first electron reaches the continuum by tunneling
ionization. As a consequence of the fact that tunneling has no classical
counterpart, this equation possesses no real solution. In the limit $%
E_{01}\rightarrow 0$, the conservation of energy for a classical particle
reaching the continuum with vanishing drift velocity is obtained. Eq. (\ref%
{saddle2}) constrains the intermediate momentum of the first electron, so
that it returns to the site of its release, which lies at the geometric
center of the molecule. In this specific case, this means the origin of the
coordinate system. Finally, Eq. (\ref{saddle3}) expresses the conservation
of energy at a later time $t$, when the first electron rescatters
inelastically with its parent ion, giving part of its kinetic energy upon
return to overcome the second ionization potential $E_{02}.$ Both electrons
then leave immediately with final momenta $\mathbf{p}_{n}.$ If we rewrite
Eq. (\ref{saddle3}) as
\begin{equation}
\sum_{n=1}^{2}[p_{n\parallel }+A(t)]^{2}=\left[ k+A(t)\right]
^{2}-2E_{02}-\sum_{n=1}^{2}\mathbf{p}_{n\perp }^{2},  \label{hypersphere}
\end{equation}
in terms of the electron momentum components parallel and perpendicular to
the laser-field polarization, this yields the equations of a hypersphere in
the momentum space, whose radius defines the region for which
electron-impact ionization is classically allowed. If the transverse
momentum components $\mathbf{p}_{n\perp }^{2}$ are kept fixed, they mainly
shift the second ionization potential towards higher values, and,
effectively decrease this region. For more details, c.f. Ref. \cite{FB2003}.

Using the above-stated saddle-point equations, the transition amplitude is
then computed by means of a uniform saddle-point approximation (see \cite%
{atiuni} for details). A more rigorous approach would be to
incorporate the prefactors (\ref{Vpnkbond}) or (\ref{Vpnkanti}) in
the action. This would lead to modified saddle-point equations, in
which the structure of the molecule, in particular scattering
processes involving one or two centers, are taken into account.
Recently, however, in the context of HHG, it has been verified that,
unless the internuclear distances are of the order of the electron
excursion amplitude, both procedures yield practically the same
results \cite{PRACL2006,F2007}. Therefore, for simplicity, we will
restrict
our investigation to single-atom saddle-point equations (\ref{saddle1})-(\ref%
{saddle3}), together with the two-center prefactors (\ref{Vpnkbond}) or (%
\ref{Vpnkanti}).

\section{Electron momentum distributions}

\label{results}

In this section, we will compute electron momentum distributions, as
functions of the momentum components $(p_{1\parallel },p_{2\parallel })$
parallel to the laser-field polarization. We approximate the external laser
field by a monochromatic wave, i.e.,
\begin{equation}
\mathbf{E}(t)=\varepsilon _{0}\sin \omega t\mathbf{e}_{x}.
\end{equation}%
This is a reasonable approximation for pulses whose duration is of the order
of ten cycles or longer (see, e.g. \cite{FLSL2004} for a more detailed
discussion). In particular, we will investigate how the symmetry of the
molecular orbitals influence the electron momentum distributions. For a
monochromatic driving field, these distributions read%
\begin{equation}
F(p_{1\parallel },p_{2\parallel })=\hspace*{-0.1cm}\iint \hspace*{-0.1cm}%
d^{2}p_{1\perp }d^{2}p_{2\perp }|M_{R}(\mathbf{p}_{n},t,t^{\prime })+M_{L}(%
\mathbf{p}_{n},t,t^{\prime })|^{2},  \label{distr}
\end{equation}%
where $M_{R}(\mathbf{p}_{n},t,t^{\prime })$ is given by Eq. (\ref{Mp}), and $%
M_{L}(\mathbf{p}_{n},t,t^{\prime })=M_{R}(-\mathbf{p}_{n},t\pm T/2,t^{\prime
}\pm T/2)$. The subscripts $\ L$ and $R$ denote the left and the right peaks
in the electron momentum distributions, respectively. Thereby, we used the
symmetry $\mathbf{A}(t)=\pm \mathbf{A}(t\pm T/2),$ and integrated over the
transverse momenta. We will also consider situations for which the
transverse momenta are resolved. In this case, the integrals in (\ref{distr}%
) are dropped.

The above-stated distribution may also be mimicked employing a classical
ensemble computation, in which a set of electrons are released with
vanishing drift momentum and weighed with the quasi-static rate
\begin{equation}
R(t^{\prime })\sim |E(t^{\prime })|^{-1}\exp \left[
-2(2|E_{01}|)^{3/2}/(3|E(t^{\prime })|)\right] .
\end{equation}%
Subsequently, these electrons propagate in the continuum following the
classical equations of motion in the absence of the binding potential.
Finally, some of them return and release a second set of electrons.
Explicitly, this distribution is given by
\begin{equation}
F^{cl}(p_{1\parallel },p_{2\parallel })=\iint \hspace*{-0.1cm}d^{2}p_{1\perp
}d^{2}p_{2\perp }F^{cl}(\mathbf{p}_{1},\mathbf{p}_{2}),
\end{equation}%
with
\begin{eqnarray}
F^{cl}(\mathbf{p}_{1},\mathbf{p}_{2})\hspace*{-0.2cm}=\hspace*{-0.2cm}
&&\int dt^{^{\prime }}R(t^{\prime })|V_{\mathbf{p}_{n}\mathbf{k}}|^{2}|V_{%
\mathbf{k}0}|^{2}  \notag \\
&&\delta \left( \sum\limits_{i=1}^{2}\frac{\left[ \mathbf{p}_{i}+\mathbf{A}%
(t)\right] ^{2}}{2}+|E_{02}|-E_{\mathrm{r}}(t)\hspace*{-0.1cm}\right) ,
\label{fclass}
\end{eqnarray}%
where $E_{\mathrm{r}}(t)=[\mathbf{k}+\mathbf{A}(t)]^{2}/2$ is the
kinetic energy of the first electron upon return (see
\cite{FSLB2004} for details). One should note that the argument in
Eq. (\ref{fclass}) is just Eq. (\ref{saddle3}), which expresses
conservation of energy following rescatter. This argument implicitly
depends on $t^{\prime },$ since both start and return times are
inter-related. If the laser-field intensity is far above the
threshold, i.e., if the classically allowed region is large, both
approaches yield very similar results \cite{thresh2006}.

\subsection{Angle-integrated distributions}

\label{gauge} As a first step, we will discuss angle-integrated
electron momentum distributions from Eq. (\ref{Mp}), for different
gauges and orbital symmetry. To first approximation, we will assume
that the second electron is dislodged by the contact-type
interaction
\begin{equation}
V_{12}(\mathbf{r}_{1}-\mathbf{r}_{2})=\delta (\mathbf{r}_{1}-\mathbf{r}_{2}),
\label{deltaee}
\end{equation}
and that the electrons are bound in $1s$ states. These assumptions have been
employed in \cite{NSDIsymm}, and led to a reasonable degree of agreement
with the experimental data. In this case, the prefactor $V_{12}(\mathbf{p}%
_{1}-\mathbf{k})=const.$ in (\ref{Vpnkbond})-(\ref{Vpnkanti}), and the
Fourier transform of the initial wave function of the second electron reads
\begin{equation}
\varphi _{0}^{(2)}(\mathbf{\mathcal{P}}(t))\sim \frac{1}{[2E_{02}+\mathbf{%
\mathcal{P}}(t)^{2}]^{2}}.  \label{psi2pdelt}
\end{equation}%
The prefactors $V_{\mathbf{k}0}$ and $V_{\mathbf{p}_{n}\mathbf{k}}$ agree
with the results in \cite{NSDIsymm}, for which the velocity gauge was taken.

We will consider the ionization potentials and equilibrium internuclear
distance of $N_{2}$, and laser-field intensities well within the
experimental range. To first approximation, we will model the
highest-occupied molecular orbital of $N_{2}$ using the symmetric prefactor (%
\ref{Vpnkbond}). In order to facilitate a direct comparison, we will also
include the antisymmetric prefactor (\ref{Vpnkanti}), and the single-atom
case, for which $V_{\mathbf{p}_{n}\mathbf{k}}\sim V_{12}(\mathbf{p}_{1}-%
\mathbf{k})\varphi _{0}^{(2)}(\mathbf{\mathcal{P}}(t))$, and employ the same
molecular and field parameters for all cases.
\begin{figure}[tbp]
\begin{center}
\includegraphics[width=9cm]{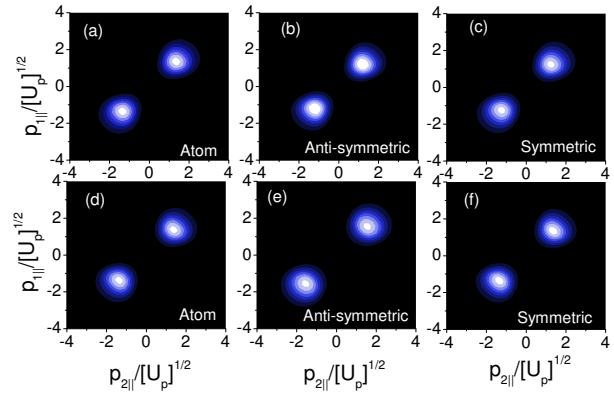}
\end{center}
\caption{Angle-integrated electron momentum distributions as functions of
the momentum components $(p_{1\parallel },p_{2\parallel })$ parallel to the
laser-field polarization, computed using the contact-type interaction (%
\protect\ref{deltaee}). The field intensity and frequency have been taken as
$I=1.5\times 10^{14}\mathrm{W/cm}^{2}$, and $\protect\omega =0.057$ a.u.,
respectively, and the ionization potentials $E_{01}=0.573$ a.u. and $%
E_{02}=0.997$ a.u.correspond to $N_{2}$ at the equilibrium
internuclear distance $R=2.068$ a.u. The upper and lower panels have
been calculated in the velocity and the length gauge, respectively.
Panels (a), and (d) correspond to the single atom case, panels (b)
and (e) to the symmetric prefactors (\protect\ref{Vk0bond}) and
(\protect\ref{Vpnkbond}), and panels
(c) and (f) to the antisymmetric prefactors (\protect\ref{Vk0anti}) and (%
\protect\ref{Vpnkanti}).}
\label{angleint}
\end{figure}

Figure \ref{angleint} depicts the above-mentioned distributions. In general,
even though different gauges and orbital symmetry lead to very distinct
prefactors, the shapes of the distributions are very similar. This is due to
the fact that the momentum region for which the transition amplitude (\ref%
{Mp}) has a classical counterpart is relatively small. Indeed, we have
verified that, for vanishing transverse momenta $p_{1\perp }=$ $p_{2\perp
}=0 $, this region starts slightly below $\pm \sqrt{U_{p}},$ and extends to
almost $\pm 3\sqrt{U_{p}}.$ This is the case for which the classically
allowed region is the most extensive, so that below $\pm \sqrt{U_{p}}$ the
contributions to the yield are negligible. Hence, the maxima and the shapes
of these distributions are determined by the interplay between phase-space
effects and the prefactor (\ref{psi2pdelt}).

In the length gauge, Eq. (\ref{psi2pdelt}) is very large near $p_{1\parallel
}+$ $p_{2\parallel }=\pm 1.5\sqrt{U_{p}},$ while in the velocity gauge this
holds for $p_{1\parallel }+$ $p_{2\parallel }=\pm 0.5\sqrt{U_{p}}$. This is
in agreement with the features displayed in Fig. 1. In fact, a closer
inspection of the length-gauge distributions shows that they exhibit
slightly larger maxima, near $p_{1\parallel }=p_{2\parallel }=\pm 1.5\sqrt{%
U_{p}}$, and are broader along $p_{1\parallel }=-p_{2\parallel }$ than their
velocity-gauge counterparts. In the velocity gauge, since the peak of the
prefactor lies outside the classically allowed region, we expect that the
yield will be maximal near the smallest momentum values which have a
classical counterpart. This agrees with Figs. \ref{angleint}(a)-(c), which
exhibit peaks slightly above $\pm \sqrt{U_{p}}$.

In Fig. \ref{angleint}, one also notices that the distributions are nearly
identical in the single-atom and molecular case. This is possibly due to the
fact that the distributions are being angle-integrated. Apart from that, we
have verified that, within the classically allowed region, there is at most
a single interference minimum. This may additionally contribute for the lack
of well-defined interference patterns.

\subsection{Interference effects}

\label{orbits}

For the above-stated reasons, in order to investigate whether interference
patterns are present in the NSDI electron momentum distributions, we will
proceed in many ways. First, we will increase the classically allowed
momentum region, and hence the radius of the hypersphere given by Eq. (\ref%
{hypersphere}). For that purpose, we will increase the intensity of the
driving laser field. Second, in this section, we will consider aligned
molecules, as it is not clear whether integrating over the alignment angle
washes the interference patterns out. One should note that, for the
parameters considered in this work, the De Broglie wavelength of the
returning electron is much larger than the equilibrium internuclear distance
of $N_{2}$.

Finally, in order to disentangle the influence of the prefactor which
accounts for the two-center interference from that of $\varphi _{0}^{(2)}(%
\mathbf{\mathcal{P}}(t)),$ we make the further assumption that $V_{12}$ is
placed at the position of the ions. Without this assumption, prefactor (\ref%
{psi2pdelt}) corresponding to the contact interaction depends on the final
electron momenta, and thus introduces a bias in the distributions. This may
obscure any effects caused solely by the molecular prefactors.

Explicitly, this reads
\begin{equation}
V_{12}=\delta (\mathbf{r}_{1}-\mathbf{r}_{2})\left[ \delta (\mathbf{r}_{2}-%
\mathbf{R}/2)+\delta (\mathbf{r}_{2}+\mathbf{R}/2)\right] .
\end{equation}%
\ Such an interaction has been successfully employed in the single-atom
case, and led to \textquotedblleft balloon-shaped\textquotedblright\
distributions peaked near $p_{1\parallel }=p_{2\parallel }=\pm 2\sqrt{U_{p}}%
. $ Such distributions exhibited a reasonable degree of agreement with the
experiments \cite{FSLB2004}. This choice of $V_{12}$ yields $\varphi
_{0}^{(2)}(\mathbf{\mathcal{P}}(t))=const$, in addition to $V_{12}(\mathbf{p}%
_{1}-\mathbf{k})=const.$ Hence, apart from effects caused by the integration
over momentum space, the shape of the distributions will be mainly
determined by the cosine or sine factor in Eqs. (\ref{Vpnkbond}) or (\ref%
{Vpnkanti}). The former and the latter case correspond to the symmetric or
antisymmetric case, respectively. The explicit interference maxima and
minima are given by Eq. (\ref{interference}).

We will now perform a more detailed analysis of such interference condition.
In terms of the momentum components $p_{i\parallel },$ or $p_{i\perp }$ $%
(i=1,2),$ parallel or perpendicular to the laser-field polarization, this
condition may be written as $\cos \left[ \zeta R/2\right] $ or $\sin \left[
\zeta R/2\right] ,$ in terms of the argument $\zeta .$ Explicitly, this
argument is given by
\begin{equation}
\zeta =\zeta _{\parallel }+\zeta _{\perp },
\end{equation}%
with
\begin{equation}
\zeta _{\parallel }=\left[ \sum\limits_{i=1}^{2}p_{i\parallel }-k(t)\right]
\cos \theta  \label{argpar}
\end{equation}%
and%
\begin{equation}
\zeta _{\perp }=p_{1\perp }\sin \theta \cos \varphi +p_{2\perp }\sin \theta
\cos (\varphi +\alpha ).  \label{argperp}
\end{equation}%
In the above-stated equations, $\theta $ gives the alignment angle of the
molecule, $\varphi $ corresponds to the angle between the perpendicular
momentum $\mathbf{p}_{1\perp }$ and the polarization plane, and $\alpha $
yields the angle between both perpendicular momentum components. Since we
are dealing with non-resolved transverse momenta, we integrate over the
latter two angles. In the velocity and in the length gauge, $k(t)=k$ and $%
k(t)=k-A(t)$, respectively. Interference extrema will then be given by the
condition%
\begin{equation}
(\zeta _{\perp }+\zeta _{\parallel })R=n\pi .  \label{interf}
\end{equation}%
For a symmetric linear combination of atomic orbitals, even and odd $n$
correspond to interference maxima and minima, respectively, whereas, in the
antisymmetric case, this condition is reversed.

An inspection of Eqs. (\ref{argpar}) and (\ref{argperp}), together with the
above-stated condition, provides an intuitive picture of how the
interference patterns change with the alignment angle $\theta .$ For
parallel alignment, the only contributions to such patterns will be due to $%
\zeta _{\parallel }$. In this particular case, the interference condition
may be written as%
\begin{equation}
p_{1\parallel }+p_{2\parallel }=\frac{n\pi }{R\cos \theta }+k(t),
\label{fringespar}
\end{equation}%
where $\cos \theta =1$. Eq. (\ref{fringespar}) implies the existence of
well-defined interference maxima or minima, which, to first approximation,
are parallel to the anti-diagonal $p_{1\parallel }=-p_{2\parallel }.$ This
is only an approximate picture, as $k$, according to the saddle-point
equation (\ref{saddle2}), is dependent on the start time $t^{\prime }$ and
on the return time $t$. Furthermore, since $t^{\prime }$ and $t$ also depend
on the transverse momenta of the electrons (see \cite{FB2003} for a more
detailed discussion), Eq. (\ref{fringespar}) is influenced by such momenta.
Finally, in the length gauge, there is an additional time dependence via the
vector potential $\mathbf{A}(t)$ at the instant of rescattering.

As the alignment angle increases, the contributions from the term $\mathbf{%
\zeta }_{\perp }$ related to the transverse momenta start to play an
increasingly important role in determining the interference conditions. The
main effect such contributions have is to weaken the fringes defined by Eq. (%
\ref{fringespar}), until, for perpendicular alignment, the fringes
completely vanish and the electron momentum distributions resemble those
obtained for a single atom. This can be readily seen if we consider the
interference condition for $\theta =\pi /2$, which is
\begin{equation}
p_{1\perp }\cos \varphi +p_{2\perp }\cos (\varphi +\alpha )=\frac{n\pi }{R}.
\label{fringesperp}
\end{equation}%
Eq. (\ref{fringesperp}) gives interference conditions which do not depend on
$k(t)$, and which vary with the angles $\varphi $ and $\alpha .$ As one
integrates over the latter parameters, which is the procedure adopted for
distributions with non-resolved transverse momentum, any structure which may
exist in Eq. (\ref{fringesperp}) is washed out.

\begin{figure}[tbp]
\begin{center}
\includegraphics[width=9cm]{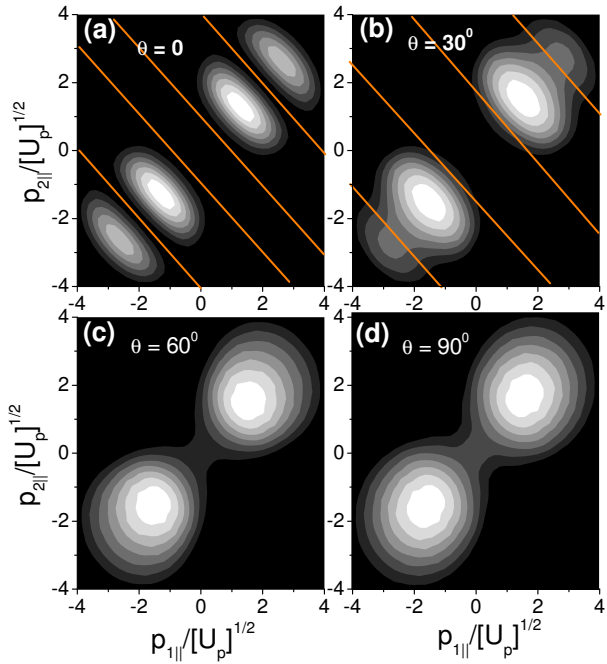}
\end{center}
\caption{Electron momentum distributions as functions of the parallel
momenta $(p_{1\parallel },p_{2\parallel })$, for several alignment angles.
We consider the velocity gauge, symmetric orbitals, and driving-field
intensity $I=5\times 10^{14}\mathrm{W/cm}^{2}$. The remaining field and
molecular parameters are the same as in the previous figure. The position of
the interference minima, estimated by assuming that the first electron
returns at a field crossing, are indicated by the lines in the figure. Panel
(a), (b), (c) and (d) correspond to alignment angles $\protect\theta =0,%
\protect\theta =30^{0},$ $\protect\theta =60^{0}$ and $\protect\theta %
=90^{0},$ respectively. }
\label{interfe1}
\end{figure}

In Fig. \ref{interfe1}, we display electron momentum distributions computed
in the velocity gauge for a symmetric highest occupied molecular orbital and
various alignment angles. The symmetric case is of particular interest,
since, recently, NSDI electron momentum distributions have been measured for
aligned $N_{2}$ molecules \cite{NSDIalign}. For parallel alignment,
interference fringes parallel to the anti-diagonal $p_{1\parallel
}=-p_{2\parallel }$ can be clearly seen, according to Eq. (\ref{fringespar}%
). For small alignment angles, such as that in Fig.
\ref{interfe1}(b), the maxima and minima start to move towards
larger parallel momenta. Furthermore, there exists an increase in
the momentum difference between consecutive maxima or minima, and
the interference fringes become less defined. This is due to the
fact that the term $\mathbf{\zeta }_{\perp }$, which washes out the
interference patterns, is getting increasingly
prominent. For large alignment angles, such as that in Fig. \ref{interfe1}%
(c), the contributions from this term are very prominent and have
practically washed out the two-center interference. Finally, for
perpendicular alignment, the distributions resemble very much those
obtained
for the single-atom case, i.e., circular distributions peaked at $%
p_{1\parallel }=p_{2\parallel }=\pm 2\sqrt{U_{p}}$ (c.f. Refs. \cite%
{FSLB2004,FL2005} for details). This is expected, since the term responsible
for the two-center interference fringes is vanishing for $\theta =90^{0}$.

The fringes in Fig. \ref{interfe1} exhibit a very good qualitative agreement
with the interference conditions derived in this section. Furthermore, the
figure shows that, for some alignment angles, the patterns caused by the
two-center interference survive the integration over the transverse momentum
components. It is not clear, however, how well the position of the fringes
agree with Eq. (\ref{fringespar}) quantitatively, and if it is possible to
provide simple estimates for these maxima and minima. Apart from that, it is
not an obvious fact that the patterns survive the integration over the
transverse momentum, and one should understand why this happens.

In particular, the role of the intermediate momentum of the first electron
will be analyzed subsequently. According to the return condition (\ref%
{saddle2}), this quantity depends on the start and return times of the first
electron. Furthermore, in the length gauge, the interference condition also
depends on the vector potential $A(t_{1})$ at the return time of the first
electron. For each pair $(p_{1\parallel },p_{2\parallel })$, the emission
and return times are strongly dependent on the transverse momenta \cite%
{FB2003}. Apart from that, physically, there are several orbits along which
the first electron may return, which occur in pairs. Hence, there exist
several possible values for $k$. In practice, only the two shortest orbits
contribute significantly to the yield. The contributions from the remaining
pairs are strongly suppressed due to wave-packet spreading. However, this
still means that the intermediate momentum, and therefore the position of
the maxima and minima, has two possible values, which depend on the start
and return times, and also on the final momentum components.

We have made a rough estimate of the position of these patterns for parallel
alignment, in the velocity and length gauges, along the diagonal $%
p_{1\parallel }=p_{2\parallel }=p_{\parallel }$. This estimate is given in
Table 1. For symmetric highest occupied molecular orbitals, the even and odd
numbers denote maxima and minima, respectively, while for antisymmetric
orbits this role is reversed. Thereby, we assumed that the first electron
left at peak field and returned at a field crossing. This gives $|k|\simeq
\sqrt{U_{p}}/(0.75\pi )$ in the saddle-point equation (\ref{saddle2}).
Furthermore, in the length-gauge estimate, we took $|A(t)|\simeq 2\sqrt{U_{p}%
}$. We have verified that both quantities are negative for the orbits in
question.

\begin{figure}[tbp]
\begin{center}
\includegraphics[width=9cm]{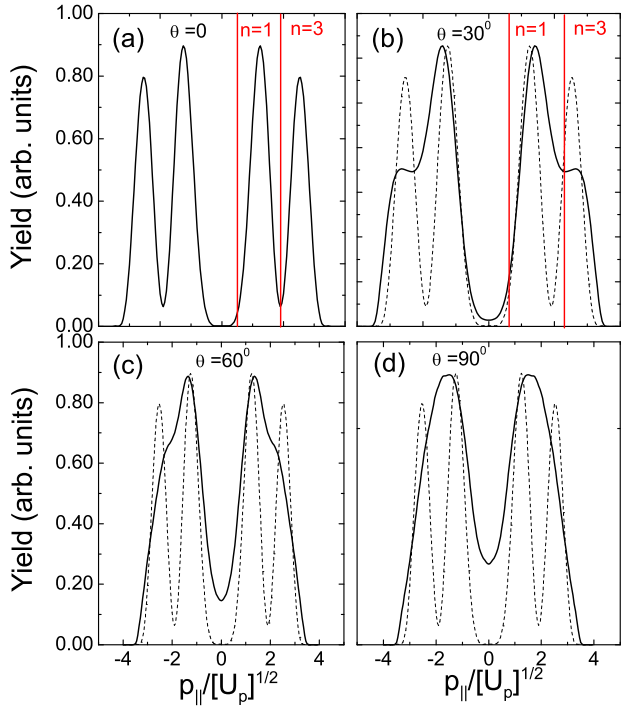}
\end{center}
\caption{Electron momentum distributions for parallel momenta $p_{1\parallel
}=p_{2\parallel }=p_{\parallel },$ non-resolved transverse momenta and
several alignment angles. We consider the velocity gauge, symmetric
orbitals, and the same molecule and field parameters as in the previous
figure. The position of the interference minima, estimated by assuming that
the first electron returns at a field crossing, are indicated by the
vertical lines in the figure. Panel (a), (b), (c) and (d) correspond to
alignment angles $\protect\theta =0,\protect\theta =30^{0},$ $\protect\theta %
=60^{0}$ and $\protect\theta =90^{0},$ respectively. For comparison, the
yield for $\protect\theta =0$ are indicated as the dashed lines in the
figure. To facilitate the comparison, the yields have been normalized to the
same peak values.}
\label{diag}
\end{figure}
These estimates agree reasonably well with the electron momentum
distributions along $p_{1\parallel }=p_{2\parallel }=p_{\parallel
}$. These distributions are depicted in Fig. \ref{diag} for several
alignment angles, the velocity gauge, and symmetric highest occupied
molecular orbitals. The positions of the minima, for each angle, are
indicated in the figure. These minima have been computed employing
Eq. (\ref{fringespar}) and the above-stated estimate for $k$. For
parallel alignment [Fig. \ref{diag}(a)], the position of the extrema
agree relatively well with Table 1. This suggests that the
intermediate momentum of the first electron, upon return, can be
approximated by its value at the field crossing. As the alignment
angle increases, the patterns become increasingly blurred until they
are eventually washed out by the contributions of $\zeta_{\perp}$.
For instance, for $\theta =30^{0}$ [Fig. \ref{diag}(b)], one may
still identify a change of slope in the distributions, at the
momentum for which the minima $n=3$ is expected to occur. For
$\theta=60^{0},$ however, the term $\zeta_{\perp}$
has already washed out the interference patterns. Indeed, in Fig. \ref{diag}%
(c), there is no evidence of interference patterns. Finally, for
perpendicular alignment, the distributions resemble very much those
obtained in the single-atom case, as shown in [Fig. \ref{diag}(d)].

\begin{center}
\begin{table}[tbp]
\begin{tabular}{c|c}
\hline\hline
\vspace*{-0.2cm} &  \\
\vspace*{-0.2cm}Extrema & Parallel momentum $p_{\parallel }/[U_{p}]^{1/2}$
\\
&  \\
\vspace*{-0.4cm} Order $n$ & \multicolumn{1}{|l}{velocity gauge \quad length
gauge} \\
&  \\ \hline
\vspace*{-0.2cm} &  \\
1 & \multicolumn{1}{|l}{0.513 \ \ \ \ \ \ \ \ \ \ \ \ \ \ \ \ \ \ \ \ \ \
1.513} \\
2 & \multicolumn{1}{|l}{1.239 \ \ \ \ \ \ \ \ \ \ \ \ \ \ \ \ \ \ \ \ \ \
2.239} \\
3 & \multicolumn{1}{|l}{1.964 \ \ \ \ \ \ \ \ \ \ \ \ \ \ \ \ \ \ \ \ \ \
2.964} \\
4 & \multicolumn{1}{|l}{2.689 \ \ \ \ \ \ \ \ \ \ \ \ \ \ \ \ \ \ \ \ \ \
3.689} \\ \hline\hline
\end{tabular}
\vspace*{0.3cm}
\caption{Electron momenta corresponding to the interference maxima and
minima given by Eq. (26), in the velocity and length gauges, for a
parallel-aligned molecule, for the same field and molecule parameters as in
Fig. 3. The parallel momenta $p_{\parallel }$ have been taken to be along
the diagonal $p_{1\parallel }=p_{2\parallel }$ in the momentum plane, and
the transverse momenta are assumed to be vanishing. If the highest occupied
molecular orbital is approximated by a symmetric combination of atomic
orbitals, the maxima and minima are denoted by even and odd number, while in
the antisymmetric case, this role is reversed, i.e., odd and even numbers
denote maxima and minima, respectively.}
\end{table}
\end{center}

\begin{figure}[tbp]
\begin{center}
\includegraphics[width=9cm]{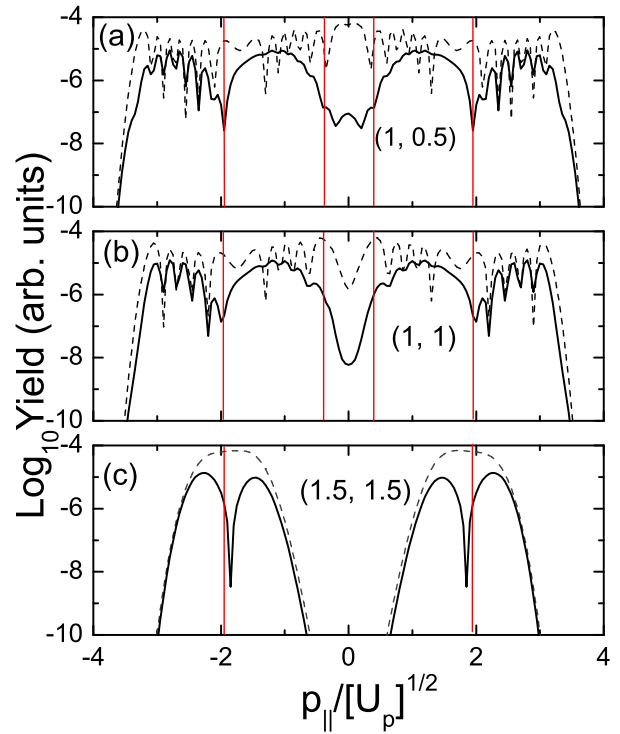}
\end{center}
\caption{Electron momentum distributions for resolved transverse
momenta, as functions of the parallel momentum $p_{1\parallel
}=p_{2\parallel }=p_{\parallel },$ for alignment angle
 $\protect\theta =0$. We consider the velocity gauge, symmetric
orbitals, and the same molecule and field parameters as in the
previous figure. For comparison, the corresponding single-atom
distributions are presented as the dashed lines in the figure. The
interference minima according to Table 1 are indicated by the
vertical lines in the figure. The numbers in the figure indicate the
transverse
momentum components $(p_{1\perp },$ $p_{2\perp })$ in units of $\protect%
\sqrt{U_{p}}.$}
\label{resolved}
\end{figure}

In order to investigate the behavior of the intermediate momentum
$k$ with respect to $p_{n\perp }(n=1,2)$, we will compute electron
momentum distributions keeping the absolute values of the transverse
momenta fixed. For simplicity, we will take $\theta =0$ and parallel
momenta along the diagonal, i.e., $p_{1\parallel }=p_{2\parallel
}=p_{\parallel }$. These distributions are displayed in Fig.
\ref{resolved}. In this case, there exists a region of parallel
momenta for which the yield is oscillating, between a maximum and a
minimum parallel momentum. These oscillations are due to the quantum
interference between the two shortest possible orbits along which
the first electron may return. These orbits constitute the pair that
has been employed in the computations performed in this work. The
larger the transverse momenta are, the less extensive this region
is. This is expected according to Eq. (\ref{hypersphere}), which
delimits this region (for details see Ref. \cite{FB2003}).

Apart from these oscillations, Fig. \ref{resolved} also exhibits the maxima
and minima caused by the spatial two-center interference. The figure shows
that the position of such patterns is very robust with respect to the choice
of $p_{\perp n}$, $n=1,2$. Indeed, both maxima and minima remain at
practically the same positions, if different transverse momenta are taken.
For this reason, such patterns survive if one integrates over the transverse
momenta. In contrast, the oscillations due to the temporal interference get
washed out. For the parameters employed in the figure, we have verified a
reasonable agreement between the second minimum and Table 1. The first
minimum is to a large extent washed out by the contributions of the events
displaced by a half-cycle, i.e., which are related to the transition
amplitude $M_L$.

\begin{figure}[tbp]
\begin{center}
\includegraphics[width=9cm]{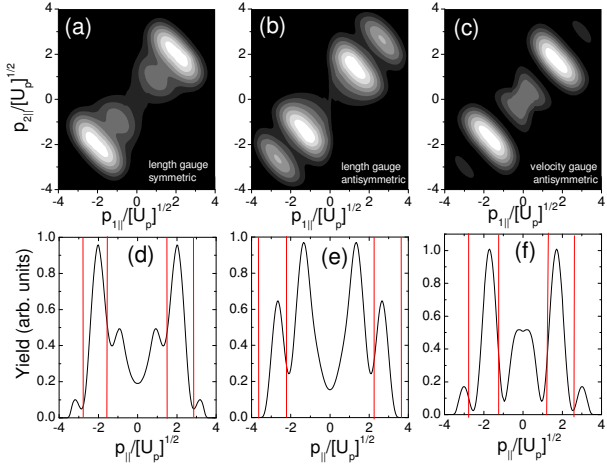}
\end{center}
\caption{Electron momentum distributions for a parallel-aligned molecule ($%
\protect\theta =0$), different orbital symmetries and gauges. The
upper and lower panels give the contour plots as functions of the
parallel momenta, and the distributions along $p_{1\parallel
}=p_{2\parallel }=p_{\parallel },$ respectively. We integrate over
the transverse momenta, and employ the same molecule and field
parameters as in the previous figures. The interference minima
according to Table 1 are indicated by the vertical lines in the
figure. Panel (a) and (d), (b) and (e), and (c) and (f) correspond
to symmetric orbitals in the length gauge, antisymmetric orbitals in
the length gauge and antisymmetric orbitals in the velocity gauge,
respectively. For panels (d), (e) and (f), the units in the vertical
axis have been chosen so that their upper values are unity (the
original values have been divided by 0.016, 0.01 and 0.04,
respectively).} \label{gaugesymm}
\end{figure}

Interference fringes parallel to $p_{1\parallel }=-p_{2\parallel }$ are also
present in the length gauge, and for antisymmetric orbitals. This is shown
in the upper panels of Fig. \ref{gaugesymm}, for parallel alignment angle.
In fact, the main difference as compared to the symmetric, velocity-gauge
case, is the position of such patterns, in agreement with Eq. (\ref%
{fringespar}). There is also some blurring in the patterns, in the length
gauge, possibly caused by the fact that the vector potential $\mathbf{A}(t)$
depends on the return time $t.$ This latter quantity is different for
different transverse momenta. The patterns, however, can be also clearly
identified in this gauge. In all cases, however, there is no evidence of a
straightforward connection between an enhancement or suppression of the
yield in the low-momentum region and the symmetry of the orbital. For
instance, in the velocity gauge, the yield is enhanced if the orbital is
antisymmetric. The length-gauge distributions, on the other hand, exhibit a
suppression in that region regardless of the orbital symmetry.

In the lower panels of Fig. \ref{gaugesymm}, we display the
distributions along \ $p_{1\parallel }=p_{2\parallel }=p_{\parallel
}$. Similarly to the velocity-gauge, symmetric case, the minima and
maxima of the distributions roughly agree with Table 1. In fact, the
even numbers in this table roughly give the position of the minima
in Figs. \ref{gaugesymm}(e) and (f), which correspond to
antisymmetric orbitals, while the odd numbers approximately yield
the minima in Fig. \ref{gaugesymm}(d), which display the
length-gauge, symmetric case. Specifically for the length-gauge
distributions [Figs. \ref{gaugesymm}(d) and (e)], there is an
overall displacement of roughly $2\sqrt{U_{p}}$ in the position of
the patterns. This is consistent with the modified interference
conditions in this case.

\subsection{The classical limit}

\label{classdistr}
\begin{figure}[tbp]
\begin{center}
\includegraphics[width=9cm]{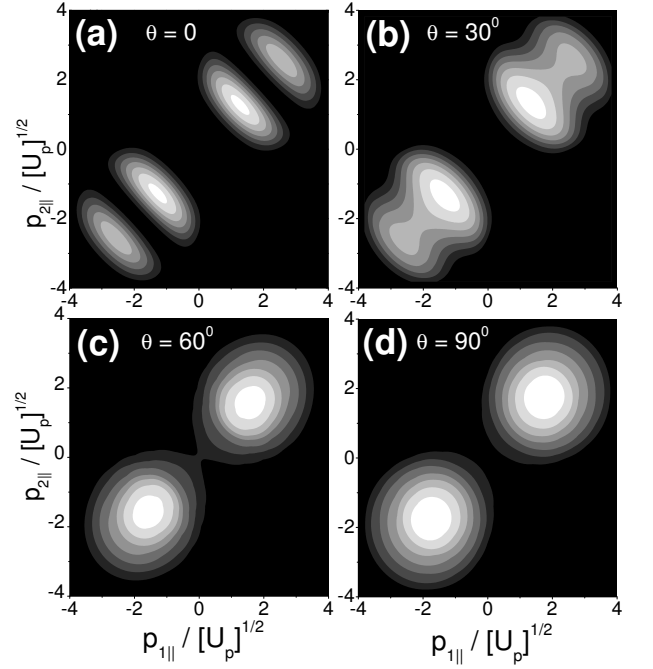}
\end{center}
\caption{Electron momentum distributions for symmetric highest occupied
molecular orbitals and several alignment angles, as functions of the
parallel momentum components $(p_{1\parallel}$, $p_{2\parallel})$, computed
in the velocity gauge using the classical model for the same field and
molecular parameters as in Fig. \protect\ref{interfe1}. Panels (a), (b), (c)
and (d) correspond to $\protect\theta=0$, $\protect\theta=30^0$, $\protect%
\theta=60^0$, and $\protect\theta=90^0$, respectively.}
\label{classical1}
\end{figure}

In the following, we perform a comparison between the S-Matrix computation
and its classical limit. In the single-atom case, both computations led to
very similar results, unless the driving-field intensity is close to the
threshold intensity \cite{thresh2006}. At this intensity, the kinetic energy
upon return is just enough to make the second electron overcome the
ionization potential. Therefore, since the intensity used in most figures is
far above the threshold intensity, one would expect similar results.

\begin{figure}[tbp]
\begin{center}
\includegraphics[width=9cm]{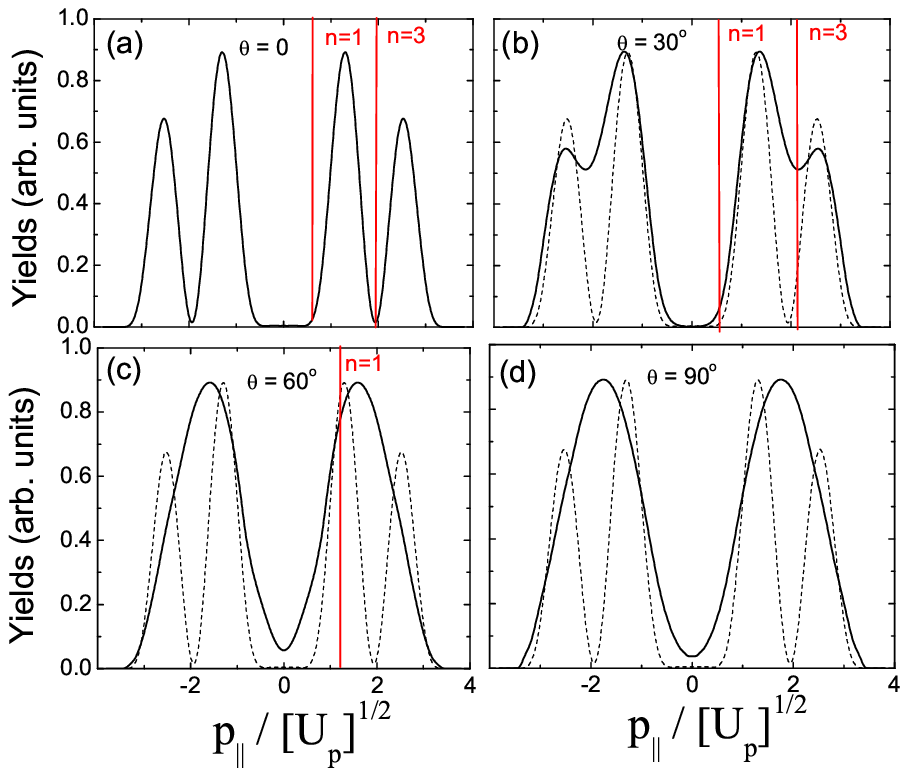}
\end{center}
\caption{Electron momentum distributions for symmetric highest occupied
molecular orbitals and several alignment angles, along $p_{1\parallel}=p_{2%
\parallel}=p_{\parallel}$, computed in the velocity gauge using the
classical model for the same field and molecular parameters as in Fig.
\protect\ref{diag}. Panels (a), (b), (c) and (d) correspond to $\protect%
\theta=0$, $\protect\theta=30^0$, $\protect\theta=60^0$, and $\protect\theta%
=90^0$, respectively. }
\label{classical2}
\end{figure}
In Fig. \ref{classical1}, we display differential momentum distributions as
functions of the parallel momentum components, computed employing the
classical model. This is the classical counterpart of Fig. \ref{interfe1},
in which the quantum mechanical distributions are depicted for the same
parameters. Indeed, for all alignment angles depicted, the classical and
quantum-mechanical distributions look very similar. Hence, even though the
two-center interference is an intrinsically quantum mechanical effect, it
can be mimicked to a very large extent within a classical model. There is
also a good quantitative agreement between the positions of the minima and
maxima in both classical and quantum mechanical cases. This is shown in Fig. %
\ref{classical2}, for parallel momenta $p_{1\parallel}=p_{2\parallel}=p_{%
\parallel}$, and several alignment angles. For $\theta=0$ and $\theta=30^0$
[Figs. \ref{classical2}(a) and \ref{classical2}(b), respectively],
the maxima and minima agree very well with those in Fig. \ref{diag}.
The main difference, with regard to the quantum-mechanical case, is
that, for large alignment angles, the classical distributions are
more localized than their quantum-mechanical counterparts,
especially in the low momentum regions. For
instance, in Fig. \ref{classical2}(d), the yield is much lower near $%
p_{\parallel}=0$, as compared to the outcome of the S-Matrix
computation [Fig. \ref{diag}(d)]. This discrepancy is possibly due
to the fact that the classical model underestimates contributions to
the yield near the boundary of the classically allowed region.

\section{Conclusions}

\label{concl}

In this work, we addressed two aspects of non-sequential double ionization
of diatomic molecules: the influence of the symmetry of the highest occupied
molecular orbital, and of the alignment angle, on the differential electron
momentum distributions. We considered the physical mechanism of
electron-impact ionization, within the strong-field approximation, and very
simple models for the highest occupied molecular orbitals, within the LCAO
and frozen nuclei approximations.

For angle-integrated electron momentum distributions, we have shown that,
for driving-field intensities within the tunneling regime and compatible
with existing experiments \cite{NSDIsymm}, the distributions computed with
symmetric and antisymmetric orbitals (prefactors (\ref{Vpnkbond}) and (\ref%
{Vpnkanti}), respectively), or different gauges, look practically identical.
This is due to the fact that, if only electron-impact ionization is taken
into account, the momentum region for which this process has a classical
counterpart is too small to allow the corresponding pre-factors to have a
significant influence. At first sight, this is in contradiction with the
experimental findings and computations in \cite{NSDIsymm}. Therein, a
broadening parallel the anti-diagonal direction has been reported only for
the anti-symmetric case, while, for a symmetric combination of atomic
orbitals, an elongation in the direction $p_{1\parallel }=p_{2\parallel }$
has been observed. One should note, however, that, in \cite{NSDIsymm}, an
effective, time-dependent second ionization potential $\ E_{02}(t)=E_{02}-2%
\sqrt{2|E(t)|}$ is used \cite{Threshold}. This feature has not been
incorporated in the present computations. It has the effect of increasing
the classically allowed momentum region and introducing an additional time
dependence in the prefactors and the action.

We have also made a detailed assessment of the interference effects due to
the fact that electron emission may occur from two spatially separated
centers. In order to disentangle the interference effects from those caused
by the prefactor $\varphi _{0}^{(2)}(\mathbf{\mathcal{P}}(t))$, we assumed
that the second electron was dislodged by a contact-type interaction at the
position of the ions. We have observed interference fringes in the electron
momentum distributions, along $p_{1\parallel }=-p_{2\parallel }+const$ for
all gauges and orbital symmetries. These fringes are most pronounced if the
molecule is aligned parallel to the laser-field polarization. As the
alignment angle increases, it gets washed out by the term (\ref{fringesperp}%
), which, for angle-integrated momenta, is essentially isotropic in the
perpendicular momentum plane. Consequently, the peaks of the distributions
shift towards higher momenta, and their shapes resemble more and more those
obtained for the same type of interaction in the single-atom case. We have
also found that the prominence of such peaks will depend on the integration
over the electron transverse momenta, so that some maxima may be more
prominent than others.

Interestingly, we are able to observe changes in the peak momenta of the
distributions, as the alignment is varied, even if a single physical
mechanism, namely electron-impact ionization, is considered. These changes
are caused by the two-center interference effects. This complements recent
results, in which different types of collisions and double-ionization
mechanisms are associated with changes in the peaks of NSDI distributions,
within the context of molecules \cite{NSDIalign,Beijing2007}. Finally, for
laser-field intensities within the tunneling regime, the distributions
obtained including only electron-impact ionization are far more localized
than those reported experimentally, and the differences between different
gauges and orbital symmetries are barely noticeable. In order to assess such
effects, it was necessary to consider much higher intensities, for which
other physical mechanisms, such as multiple electron recollisions, would
also be expected to play a role \cite{Beijing2007}. These discrepancies may
be due to the fact that we are not including the physical mechanism in which
the first electron, upon return, promotes the second electron to an excited
state, from which it subsequently tunnels out.

\noindent \textbf{Acknowledgements:} W.Y. would like to thank the University
College London for its kind hospitality. This work has been financed by the
UK EPSRC (Grant no. EP/D07309X/1) and by the Faculty Research Fund of the
School of Mathematics and Physical Sciences (UCL). We also thank the UK
EPSRC for the provision of a DTA studentship.

\end{document}